\documentclass[journal]{IEEEtran}
\usepackage{cite}

 \usepackage{graphicx}%
\usepackage{multirow}%
\usepackage{amsmath,amssymb,amsfonts}%
\usepackage{amsthm}%
\usepackage{mathrsfs}%
\usepackage{xcolor}%
\usepackage{textcomp}%
\usepackage{manyfoot}%
\usepackage{booktabs}%
\usepackage{algorithm}%
\usepackage{algorithmicx}%
\usepackage{algpseudocode}%
\usepackage{listings}
\usepackage{marginnote}
\usepackage{enumitem}
\usepackage{mdwlist}
\usepackage[font=footnotesize,labelformat=parens]{subcaption}
\usepackage[font=footnotesize,labelfont=bf]{caption}
\usepackage{changes}
\usepackage[table]{xcolor}
\definecolor{ampone}{RGB}{0,90,160}   
\definecolor{amptwo}{RGB}{140,70,0}   
\definecolor{ampthree}{RGB}{0,0,0}   

\ifCLASSINFOpdf
\else
\fi
\begin{document}
%
\title{The Anatomy of RF Chains: Metrics, Measures,
and Operating Efficiency}
%
%
%

\author{Leonid~Belostotski,~\IEEEmembership{Senior Member,~IEEE,}
        Arjuna~Madanayake,~\IEEEmembership{Senior Member,~IEEE,}
        John~Nielsen, Mostafa~Abdelhadi, Xingchen~Liu, Xinquan~Wang, Guanyue~Qian, Sabit~Ekin,~\IEEEmembership{Senior Member,~IEEE,}
        and~Theodore~S.~Rappaport,~\IEEEmembership{Fellow,~IEEE}
}

\maketitle

\newif\ifshowchanges
\showchangesfalse 

\ifshowchanges
\else
  \setdeletedmarkup{}
  \setaddedmarkup{\textcolor{black}{#1}}
  \definechangesauthor[color=black]{LB}
\fi

\begin{abstract}
In 1958, Haus and Adler~\cite{Haus1958} introduced the concept of \textit{noise measure}. Noise measure is a single quantitative metric that provides a comprehensive basis for comparing devices (individual circuits or outcomes of optimization iterations) in terms of their contribution to overall system noise by incorporating both noise factor and available power gain. Unlike noise factor alone, which reflects how much a device degrades the signal-to-noise ratio, noise measure captures the trade-off between noise and gain, making it a \textit{system-aware} metric. This distinction is especially important when comparing devices in multistage systems, where both parameters jointly influence the overall system noise.
Building on Haus and Adler’s work, this article aims to advance RF system design by extending traditional device metrics, such as noise factor and noise measure, with new system-aware metrics: linearity, dynamic range, power efficiency, and waste measures. These new measures are interpretable, computable, and cascadable, making them well-suited for comparing the impact of individual devices or tracking the convergence of circuit design iterations on overall system-level performance.  Additionally, a new metric---\textit{operating efficiency}---is introduced, which unifies power efficiency, dynamic range, and data rate by incorporating signal statistics and variability in communication circuits and systems. Operating efficiency enables robust evaluation of devices under realistic and transient operating conditions, including interference, modulated signals, and adaptive modulation schemes.
\end{abstract}


%
\IEEEpeerreviewmaketitle
\bstctlcite{IEEEexample:BSTcontrol}

\section{Introduction}

Modern RF communication systems face competing demands on signal quality, linearity, dynamic range, power consumption, and implementation complexity. Traditional device-level metrics, such as noise factor, intercept points, compression points, and power efficiency, are indispensable for characterizing isolated components but provide limited guidance for system-level reasoning or for comparing architectural choices in cascaded RF subsystems.

As systems scale in bandwidth, data rate, and architectural complexity, designers increasingly require descriptions that relate device-level properties to system-level outcomes such as error vector magnitude (EVM), bit error rate (BER), achievable data rate, and total power efficiency. Reliance on individual metrics alone can yield misleading or trivial optima, motivating the need for performance descriptors that embed system context and physical trade-offs.

This article discusses a conceptual, system-aware framework for organizing RF performance descriptors and clarifying how device characteristics collectively influence cascaded system behavior. The framework builds on the classical noise measure of Haus and Adler~\cite{Haus1958} and extends the notion of \emph{measures}—quantities that combine multiple device metrics to reflect their joint impact on system performance. New measures for linearity, dynamic range, and power efficiency are presented. These measures emphasize physical interpretation, cascading behavior, and avoidance of trivial optimization outcomes. 

Within this framework, the paper also introduces \emph{operating efficiency}, a system-level descriptor linking power efficiency, dynamic range, and data rate through the statistical properties of communication signals. Operating efficiency captures practical trade-offs due to modulation, interference, and signal-envelope variability, which complements existing metrics by explicitly connecting signal-quality requirements to power consumption. Unlike prior work that treats efficiency and signal quality separately, operating efficiency provides a framework that links noise, linearity, dynamic range, and efficiency.

\section{\protect\label{sec:Signal-quality}Signal Quality}

In this section, we use a simplified and illustrative framework to discuss how physical-layer signal-quality metrics, particularly BER and EVM, relate to achievable data rate. Specifically, a framework based on M-QAM modulation over an additive white Gaussian noise (AWGN) channel is used to clarify these relationships and to connect device-level behavior to system-level performance constraints. This framework is not required for the derivation of the proposed \textit{measures}, but serves to illustrate how hardware imperfections influence communication performance. This simplified framework also provides an intuitive link to Shannon capacity and has been used previously in the context of energy-efficiency analysis~\cite{Rappaport_2011,Rappaport2025}. More comprehensive system-level treatments, such as~\cite{Bjornson2017}, also establish these connections by explicitly accounting for transceiver hardware impairments, interference, and their impact on channel capacity and overall network performance. Here, however, we adopt the simplified model to focus on device performance metrics and to highlight how hardware imperfections influence BER, EVM, and ultimately achievable data rate, without the added complexity of detailed channel and interference modeling.

The maximum theoretical data rate of an AWGN channel is given by Shannon’s capacity:
\begin{equation}
R = B \log_{2}\left ( 1+\text{SNR}_0\right )=B \log_{2}\left(1+\text{SNR}/\Gamma \right),
\label{eq:max-data-rate-1}
\end{equation}
where $B$ is the signal bandwidth and $\text{SNR}_0$ is the minimum signal-to-noise ratio (SNR) required to achieve capacity under ideal transmission and reception. Practical systems, however, must operate at higher SNR due to non‑ideal modulation, coding, synchronization, and circuit impairments. These deviations are captured through the \emph{SNR gap} $\Gamma$, 
where $\Gamma \ge 1$ quantifies the implementation loss relative to the Shannon limit~\cite{Cioffi_SNRgap,Garcia-Armada_SNRgap,Fung_SNRgap}. A larger $\Gamma$ indicates a greater required SNR for a target data rate.
Although simplified, the M‑QAM/AWGN model is useful because it makes explicit the relationships among system‑level objectives (data rate), signal‑quality constraints (BER and EVM), and device‑level limitations such as noise, nonlinearity, and dynamic range. As shown later in the article, these connections also underpin the definitions of operating dynamic range and operating efficiency. By tying SNR and BER requirements to EVM, which directly reflects device impairments, the model provides a convenient bridge between traditional device metrics and the system‑aware measures introduced in subsequent sections.

\subsection{Signal Quality Metrics: BER and EVM}
\begin{figure}[]
    \centering
        \centering
        \includegraphics[width=0.75\columnwidth]{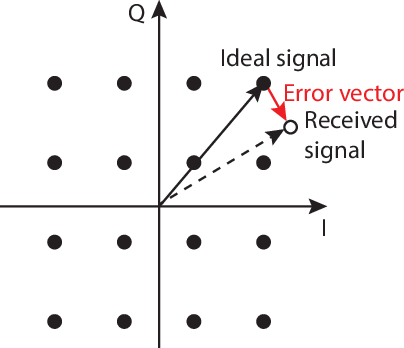}
        \label{fig:Error-vector-magnitude}
        
    \caption{Definition of the error vector magnitude (EVM) on the I/Q plane. For each transmitted symbol, the error vector connects the ideal constellation point to the actually received point; EVM is the rms length of this error vector, normalized to the reference signal magnitude, and captures the combined effect of noise, distortion, phase noise, and I/Q imbalance.}
    \label{fig:EVM_vector}
\end{figure}

In digital communication systems, link quality is commonly quantified using BER, defined as the ratio of incorrectly received bits to the total number of transmitted bits over a given interval~\cite{Rappaport1991}. BER is valuable for end‑to‑end reliability assessment, but because BER reflects the combined influence of the channel, modulation, coding, synchronization, and hardware impairments, BER offers limited diagnostic insight into individual RF components.
EVM, by contrast, is well suited for isolating device‑level impairments. EVM measures the deviation between the received and ideal constellation points (see Fig.~\ref{fig:EVM_vector})~\cite{Plett_book,Bjornson2017}.

Because EVM directly captures waveform distortion—such as noise, phase noise, I/Q imbalance, gain compression, and nonlinearities—it enables rapid assessment and correlates well with device behavior across modulation formats~\cite{RappaportBook}. Consequently, modern wireless standards (Wi‑Fi, LTE, 5G) specify EVM limits rather than BER thresholds.

Although BER and EVM quantify signal quality in different ways, they are closely related. Larger EVM is due to larger symbol scatter and therefore a higher likelihood of bit errors. For M‑QAM modulation in an AWGN channel with hard‑decision detection, this relationship may be approximated as~\cite{Goldsmith2005}
\begin{equation}
\text{BER} \approx Q\left(\frac{m}{\text{EVM}}\right),
\label{eq:BER-defined}
\end{equation}
where $Q(\cdot)$ is the Gaussian tail probability and $m = \sqrt{3/(M-1)}$ for M‑QAM.

\begin{figure}[]
    \centering
    
        \centering
        \includegraphics[width=\columnwidth]{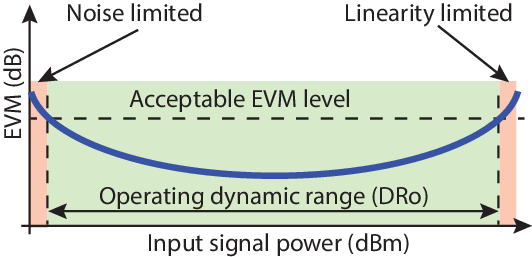}
    \caption{Typical EVM of an RF device as a function of input signal power.}
    \label{fig:EVM}
\end{figure}

Fig.~\ref{fig:EVM}
illustrates how device behavior shapes EVM as a function of input power. At low power, noise dominates ($\text{EVM}_\text{noise}$); at high power, nonlinear distortion dominates ($\text{EVM}_\text{linearity}$). Additional impairments, such as phase noise or I/Q mismatch, dominate in the intermediate region~\cite{Georgiadis2004}. 
Because noise and distortion dominate at opposite ends of the input‑power range, EVM naturally defines the operating dynamic range.

\subsection{\protect\label{subsec:Dynamic-range}Dynamic Range}

The operating dynamic range  
\begin{figure}[]
    \centering
        \includegraphics[width=\columnwidth]{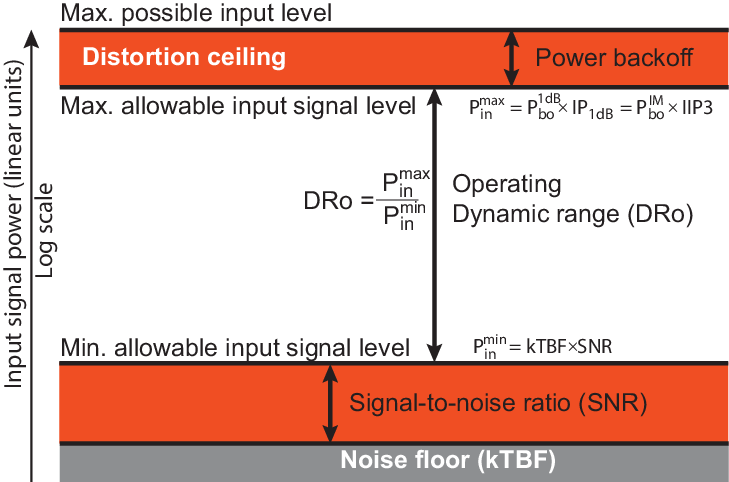} 
    \caption{Operating dynamic range of an active RF component or system, defined as the region over which the component or system maintains acceptable signal quality.}
    \label{fig:Dynamic_range}
\end{figure}
of a device is illustrated and defined in Fig.~\ref{fig:Dynamic_range} as $\text{DRo}=P_\text{in}^\text{max}/P_\text{in}^\text{min}$,
where $P_\text{in}^\text{max}$ is the maximum input signal level that can be applied before excessive distortion occurs, and $P_\text{in}^\text{min}=kTBF\times SNR$ is the minimum input power required to maintain a sufficient SNR for successful signal reception. Here, $k$ is Boltzmann's constant, $T$ is the absolute temperature, $B$ is the bandwidth, and $F$ is the noise factor. Both $P_\text{in}^\text{min}$ and $P_\text{in}^\text{max}$ are linked to the acceptable BER and EVM, thus determining the required SNR and power backoff (Pbo)---relative to the maximum possible input level, as discussed next---necessary for achieving acceptable device linearity.

There are two common ways to express $P_\text{in}^\text{max}$. The first way is relative to the power level of the 1-dB compression point (P\textsubscript{1dB}), which is the power level at which the gain of a circuit reduces by 1\,dB relative to the gain at low-power levels. The second way is relative to the power level at which intermodulation components---appearing from nonlinear behavior under strong input signals---remain below a level that violates the desired EVM specification. This type of linearity metric is typically quantified using the 3rd-order intercept point (IP3)—or, in wideband or low-IF systems where even-order distortion dominates, the 2nd-order intercept point (IP2)—with the resulting dynamic range often referred to as the spurious-free dynamic range (SFDR). P\textsubscript{1dB} and IP3 can be either input-referred (IP\textsubscript{1dB} and IIP3) or output-referred (OP\textsubscript{1dB} and OIP3). In terms of input-referred parameters, $P_\text{in}^\text{max}$ is defined in Fig.~\ref{fig:Dynamic_range},  

where $P_\text{bo}^\text{1dB}$ and $P_\text{bo}^\text{IM}$ represent the power backoff factors from the linearity metrics denoted by IP\textsubscript{1dB} and IIP3, respectively. 

\begin{figure}[]
        \centering
        \includegraphics[width=\columnwidth]{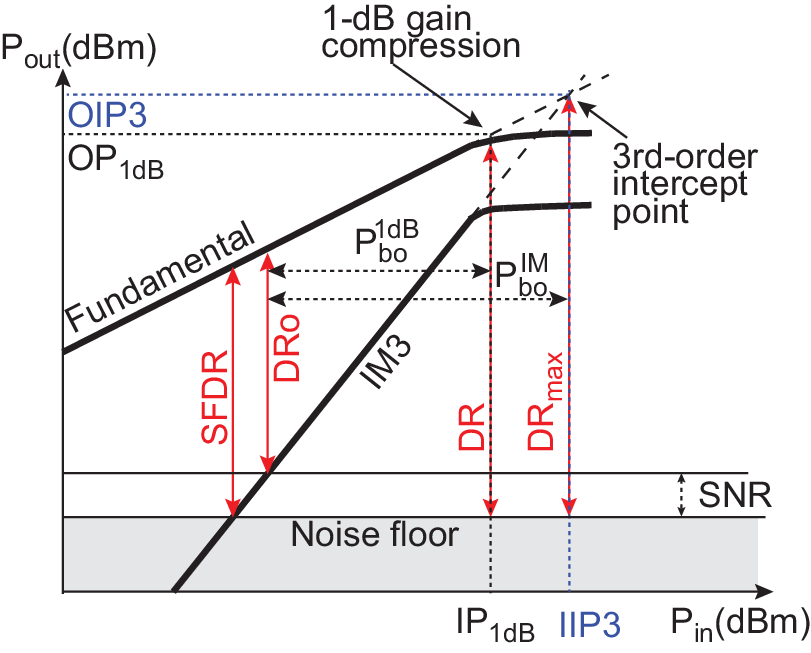}
    \caption{Linearity and dynamic-range metrics on the $P_\text{out}$–$P_\text{in}$ plane. The slope-1 fundamental and slope-3 IM3 line intersect at the third-order intercept point (IIP3, OIP3); the 1-dB compression point ($\text{IP}_\text{1dB}$, $\text{OP}_\text{1dB}$) lies below it. $P_\text{bo}^\text{1dB}$ and $P_\text{bo}^\text{IM}$ are the power backoffs from IIP3 and OIP3 defining the upper edge of $\text{DRo}$. Four dynamic-range definitions are shown, all of the form $P_\text{in}^\text{max} / P_\text{in}^\text{min}$: $\text{DR}_\text{max}$ (IIP3 over $kTBF$), SFDR (IM3-limited upper edge over $kTBF$), DR ($\text{IP}_\text{1dB}$ over $kTBF$), and $\text{DRo}$ ($\text{IIP3}\cdot P_\text{bo}^\text{IM}$ over $kTBF\cdot \text{SNR}$, the EVM-bounded operating range used in this article).}
    \label{fig:Dynamic_range_metrics}
\end{figure}

Fig.~\ref{fig:Dynamic_range_metrics}  
illustrates the key linearity metrics along with different definitions of dynamic range, including the maximum dynamic range (DR\textsubscript{max}), defined as $\text{DR}_\text{max}\equiv \text{IIP3}/kTBF$, where $P_\text{noise}=kTBF$ represents the noise floor of a device.

\subsection{\protect\label{subsec:Operating-DR}Operating Dynamic Range (DRo) and EVM}

The selection of an appropriate DRo is a system-level design decision that balances noise and linearity to ensure signal integrity under practical operating conditions. In a communication system, the transmitter generates a signal that meets a specified EVM limit, largely set by transmitter nonlinearity. As the signal propagates through the channel, AWGN and interference further degrade EVM toward the maximum level tolerated by the receiver. While signal processing and filtering suppress much of this interference, strong in-band interferers can still produce intermodulation distortion, requiring additional dynamic range to prevent excessive EVM degradation. In this context, DRo quantifies the maximum signal or interference power for which intermodulation distortion remains sufficiently low to meet EVM requirements.

The relationship between the noise-dominated EVM (expressed as a linear ratio throughout) and SNR is \cite{Steer_ARFTG2004,Mahmoud2009,Santos2007,Georgiadis2004}
\begin{equation}
\text{EVM}_{\text{noise}}\approx\text{SNR}^{-1/2},\label{eq:EVM-SNR}
\end{equation}
where an AWGN channel and ideal demodulation are assumed. 
Using \eqref{eq:EVM-SNR} and $P_\text{in}^\text{min}$ (see Fig.~\ref{fig:Dynamic_range}), the minimum input power required to maintain a target SNR is $P_\text{in}^\text{min}\approx kTBF/\text{EVM}_{\text{noise}}^{2}$.

EVM is also affected by nonlinear distortion induced by strong input signals. Its impact is captured by the ratio of input-referred third-order intermodulation power ($P_\text{IM}$) to input signal power: $\text{EVM}_{\text{linearity}}=\sqrt{P_\text{IM}/P_\text{in}}$ \cite{Plett_book}, where $P_\text{IM}=P_\text{in}^{3}/\text{IIP3}^{2}$ \cite{Sansen1999}. At the maximum input level, this gives $P_\text{in}^\text{max}\approx\text{EVM}_{\text{linearity}}\times \text{IIP3}$.

$P_\text{in}^\text{min}$ and $P_\text{in}^\text{max}$ define DRo in Fig.~\ref{fig:Dynamic_range} and the dynamic-range efficiency metric $\eta_{\text{DR}}=\text{DRo}/\text{DR}_\text{max} \approx \text{EVM}_{\text{linearity}} \times \text{EVM}_{\text{noise}}^2$. Since RF devices are typically designed to operate within DRo under equal-EVM conditions (see Fig.~\ref{fig:EVM}), substituting $\text{EVM} = \text{EVM}_{\text{noise}} = \text{EVM}_{\text{linearity}}$ gives $\eta_{\text{DR}} \approx \text{EVM}^3$. Thus, more stringent EVM requirements directly reduce DRo. This metric establishes an intuitive link between dynamic range and allowable EVM, and is consistent with measured data reported in~\cite{Nauta2020}.

\subsection{\protect\label{subsec:Data-rate}Data Rate and Dynamic Range}

To relate data rate to dynamic range, we consider the simplified AWGN channel
and M-QAM modulation introduced in Section~\ref{sec:Signal-quality}, together
with the capacity expression in~\eqref{eq:max-data-rate-1}. The use of channel capacity as the performance measure is
justified for arbitrary input distributions by the results
of~\cite{Guo2005}; the M-QAM-specific expressions that follow
are a practical specialisation of this general bound. Substituting the SNR–EVM relationship from \eqref{eq:EVM-SNR} into \eqref{eq:max-data-rate-1} and 
$\eta_{\text{DR}} \approx \text{EVM}^3$ yields the dynamic-range efficiency $\eta_\text{DR}$ in terms of data rate
\begin{align}
\eta_\text{DR} =\left[\Gamma\left(2^{R/B}-1\right)\right]^{-3/2}.
\label{eq:Data-rate-F-DR-IP-1}
\end{align}

Equation \eqref{eq:Data-rate-F-DR-IP-1} shows that increasing data rate requires a reduction in $\eta_\text{DR}$ as discussed in Sidebar~A. 

\section{\protect\label{sec:Performance-Factors}Dynamic-Range Determinants: Metrics and Measures}

Section~\ref{sec:Signal-quality} examined the relationship between data rate and signal quality, specifically EVM and BER, through the concept of dynamic range. In this section, we discuss the key traditional noise and linearity metrics---noise factor, P\textsubscript{1dB}, and IP3---that define the dynamic range of active components. We also introduce new system-aware noise, linearity, and dynamic range measures that offer a more meaningful basis for comparing and optimizing active devices than the traditional metrics.

\subsection{\protect\label{subsec:Noise-Factor}Noise Factor and Noise Measure}

The amount of noise generated by a device can be quantified using the noise factor ($F$), which is a device performance metric that is defined as a ratio of the SNRs at the input and output of the device $F=\text{SNR}_{\text{in}}/\text{SNR}_{\text{out}}$.

\begin{figure*}
\centering
\begin{centering}
\includegraphics[width=1\textwidth]{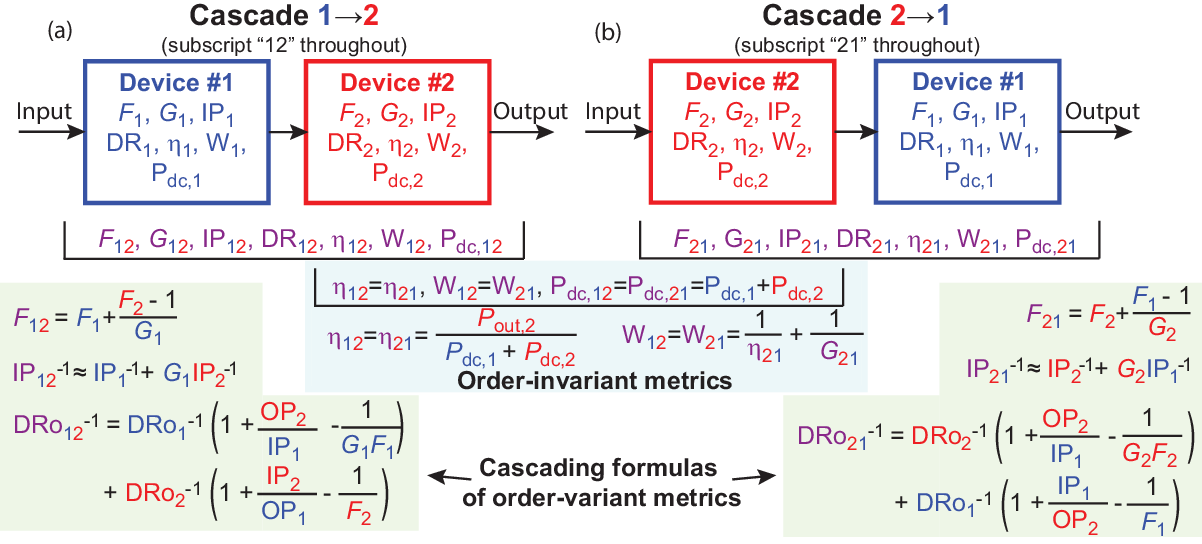}
\par\end{centering}
\caption{\protect\label{fig:Two-possible-cascaded}Two possible cascades of two devices: (a) Device \#1 followed by Device \#2; (b) Device \#2 followed by Device \#1. Subscripts “\textcolor{blue}{1}” and “\textcolor{red}{2}” denote metrics and measures associated with Device \#1 and Device \#2, respectively, while “\textcolor{blue}{1}\textcolor{red}{2}” and “\textcolor{red}{2}\textcolor{blue}{1}” correspond to the cascades in (a) and (b). Noise factor ($F$) is discussed in Subsection~\ref{subsec:Noise-Factor}; linearity ($\text{IP}$) in Subsection~\ref{subsec:Linearity-measure}; dynamic range (DR) in Subsection~\ref{subsec:Dynamic-Range}; power efficiency ($\eta$) and waste factor ($W$) in Section~\ref{sec:System-Efficiency}. Device and cascade gains ($G$), dc power consumptions ($P_\text{dc}$), cascaded $F$, IP, DRo, $\eta$, and $W$ are also indicated. Note that converting $\text{DRo}_\text{12}$ and $\text{DRo}_\text{21}$ to DR, $\text{DR}_\text{max}$, or SFDR simply involves replacing DRo with the desired metric.  The cascading formulas define the measures introduced in this article: $M_\text{N}$ (Subsection~\ref{subsec:Noise-Factor}), $M_\text{L}$ (Subsection~\ref{subsec:Linearity-measure}), $M_\text{D}$ (Subsection~\ref{subsec:Dynamic-Range}), and $M_\eta$ and $M_\text{W}$ (Section~\ref{sec:System-Efficiency}).}
\end{figure*}

The noise factor can be used to determine the device noise power level (i.e., the noise floor) via $P_{\text{noise}}=kTB\times F$, as was previously discussed in connection with $\text{DRo}$ and $P_\text{in}^\text{min}$, as well as in the definition of $\text{DR}_\text{max}$ in Subsection~\ref{subsec:Dynamic-range}. The noise factor $F_{12}$ of a cascade of two devices (Device \#1 and Device \#2) is found by applying Friis' formula \cite{Friis1944} as shown in Fig.~\ref{fig:Two-possible-cascaded}(a)
where $G$ represents the gain, and the subscripts denote the device within the cascade.

Noise factor $F$ is an important metric for identifying the sensitivity and SNR of a system, and, as such, noise factor is routinely used during device and system design. However, as Friis' formula indicates, available power gain ($G$) is also important and should be considered when comparing or optimizing devices. 

To apply Friis' formula when optimizing Device~\#1, the noise factor of Device~\#2 must be known, which is often not the case during early design. If Device~\#1 is instead optimized solely by minimizing its own noise factor $F_1$, independent of $F_2$, the design is driven toward the ideal $F_1=1$, corresponding to a through connection with unity gain ($G_1=1$). While this achieves the minimum possible noise factor, it provides no system-level benefit, since the cascaded noise factor reduces to $F_{12}=F_2$. Thus, an extremely low $F_1$ does not necessarily improve overall system performance. The same issue arises if Device~\#2 is optimized in isolation.

Haus and Adler's \cite{Haus1958} investigation of ways to answer these questions and account for the trade-off between $F$ and $G$ led to the introduction of \textit{noise measure} $M_\text{N}\overset{\triangle}{=}(F-1)/(1-G^{-1})$. To understand the origin of noise measure, consider two possible cascades of two devices as shown in Fig.~\ref{fig:Two-possible-cascaded}.
These two cascades realize two different noise factors, $F_\text{12}$
and $F_\text{21}$ via Friis' formulas in Fig.~\ref{fig:Two-possible-cascaded} where $G_\text{1}$ and $G_\text{2}$ are the available power gains of the two devices. Assuming $F_\text{12}<F_\text{21}$ yields
\begin{equation}
\overbrace{\frac{F_\text{1}-1}{1-G_\text{1}^{-1}}}^{M_{\text{N,}1}}<\overbrace{\frac{F_\text{2}-1}{1-G_\text{2}^{-1}}}^{M_{\text{N,}2}},
\end{equation}
which demonstrates that \textit{the lowest noise factor for a cascade is achieved by positioning the device from input to output in increasing order of} $M_\text{N}$
\textit{rather than $F$.} $M_\text{N}$ is the \textit{noise measure}.

Haus and Adler \cite{Haus1958} also specified several important fundamental properties of $M_\text{N}$ that can assist designers in assessing the noise performance of cascaded devices. Although these properties are outside the scope of this article, readers interested in receiver and low-noise-amplifier (LNA) design are encouraged to review \cite{Haus1958} for more detail.

The noise-measure derivation proposed by Haus and Adler motivated the other \textit{measures} presented in this article.

\subsection{\protect\label{subsec:Linearity-measure}Linearity and Linearity Measure}

Device linearity is commonly quantified by $\text{IP}_{\text{1dB}}$, IIP3, or their output-referred counterparts (see Subsections~\ref{subsec:Dynamic-range} and~\ref{subsec:Operating-DR}). These metrics estimate distortion under large-signal excitation and relate to the maximum input power $P_\text{in}^\text{max}$ as in Fig.~\ref{fig:Dynamic_range}. For a cascade, $\text{IP}_\text{12}$ linearity combines approximately as in Fig.~\ref{fig:Two-possible-cascaded} \cite{LeeTH_1}, where $\text{IP}_{12}$ denotes either IP\textsubscript{1dB} or IIP3 of the cascade.

Optimizing devices solely by IP\textsubscript{1dB} or IIP3 can lead to trivial through-connection solutions, similar to the issue with noise factor. To avoid this, Haus and Adler’s approach inspired the \textit{linearity measure}~\cite{LNA_FOM}
\begin{equation}
M_L \triangleq \text{IP} \times (G - 1),\label{eq:Linearity-measure}
\end{equation}
derived from comparing two cascade orders $\text{IP}_{12}$ and $\text{IP}_{21}$ in Fig.~\ref{fig:Two-possible-cascaded}.
When $\text{IP}_{12} > \text{IP}_{21}$, the condition $\text{IP}_1(G_1 - 1) < \text{IP}_2(G_2 - 1)$ leads to \eqref{eq:Linearity-measure}. Therefore, \textit{devices should be arranged in increasing $M_L$ to maximize system linearity.}

\subsection{Dynamic Range and Dynamic-Range Measure\protect\label{subsec:Dynamic-Range}}

As reviewed in Subsection~\ref{subsec:Dynamic-range} and Fig.~\ref{fig:Dynamic_range_metrics}, common dynamic-range definitions (DR, DRo, $\text{DR}_\text{max}$, SFDR) all take the form $P_\text{in}^\text{max}/P_\text{in}^\text{min}$, differing only in how these limits are defined relative to the fundamental response ($P_\text{in}^\text{max} = \text{IP} \times P_\text{bo}$) and the noise floor ($P_\text{in}^\text{min} = kTBF \times \text{SNR}$). To avoid repeating the same derivations for the different dynamic-range definitions, we use DRo in the remainder of this subsection.

Having identified noise and linearity as the fundamental determinants of dynamic range in Subsections~\ref{subsec:Noise-Factor} and~\ref{subsec:Linearity-measure}, we now examine how the dynamic ranges of individual devices combine in a cascade. We define the cascaded DRo, which, to the best of our knowledge, has not been explicitly reported in the literature.

From its definition, $\text{DRo}_\text{12}=P_\text{in,12}^\text{max}/P_\text{in,12}^\text{min}$ for Cascade~1$\to$2 in Fig.~\ref{fig:Two-possible-cascaded}.
Special cases arise when the cascaded DRo is limited purely by noise or purely by linearity. These cases are summarized in Sidebar~B.

In the general case, $P_\text{in,12}^\text{max}$ is determined by the cascade linearity, while $P_\text{in,12}^\text{min}$ is set by the cascade noise floor. Using $F_{12}$ and $\text{IP}_\text{12}$ from Fig.~\ref{fig:Two-possible-cascaded} to express the operating dynamic range yields $\text{DRo}_\text{12}$ that is shown in Fig.~\ref{fig:Two-possible-cascaded}, where $\text{OP}_{1(2)}=G_{1(2)}\text{IP}_{1(2)}$ denote the output-referred linearity metrics for Devices \#1 and \#2, and also yields 
\begin{equation}
\overbrace{\left(M_\text{N1}+\frac{G_\text{1}G_\text{2}}{G_\text{1}G_\text{2}-1}
\right)M_\text{L1}}^{M_\text{D,1}}<
\overbrace{\left(M_\text{N2}+\frac{G_\text{1}G_\text{2}}{G_\text{1}G_\text{2}-1}
\right)M_\text{L2}}^{M_\text{D,2}}
\label{eq:DR-measure-derivation}
\end{equation}
where we assume that $\text{DRo}_\text{12}>\text{DRo}_\text{21}$, i.e., Cascade 1$\to$2 in Fig.~\ref{fig:Two-possible-cascaded} is assumed to exhibit a higher DRo.

Expression~\eqref{eq:DR-measure-derivation} defines \textit{dynamic-range measures} $M_\text{D,1}$ and $M_\text{D,2}$ of the two cascades.

Unlike $M_N$ and $M_L$, which are fully device-intrinsic, $M_\text{D}$ retains a mild dependence on the gain of the cascade through the gain term $G_1G_2/(G_1G_2-1)$. For $G_1G_2 \gg 1$, this term approaches 1, and $M_\text{D} \approx (M_\text{N}+1)M_\text{L} \approx F \times \text{OP}$, depending only on each device's own $F$ and $\text{OP}$. Thus, $M_\text{D}$ is effectively device-intrinsic in most practical systems.

Two conclusions are drawn from the discussions in this subsection: 
1)~$\text{DR}\propto\text{OP}/F$ (a ratio) while $M_\text{D}\propto F\times\text{OP}$ (a product), so optimizing $M_\text{D}$ is not the same as optimizing DR directly but a device-level proxy that enables cascade ordering and device comparison; and 
2)~ascending $M_\text{D}$ ordering selects the DR-optimal ordering of pairs of devices.

A corollary of~1) is that no fixed ordering based solely on $M_\text{D}$ can, in general, maximize DR of more than two devices because of conflicting demands from $F$ and $\text{OP}$, implying the need for context-aware adaptive ordering. To see why, treat the first two devices $A$ and $B$ as a single combined stage and apply the two-device $M_\text{D}$ ascending ordering to the pair (AB,~C): correct ordering requires $M_\text{D,AB}<M_\text{D,C}$. In the high-gain limit this simplifies to $M_\text{L,B}(M_\text{N,A}+1)<M_\text{D,C}$, showing that the effective linearity of $B$ is weighted by the noise of the preceding stage $A$. In other words, the ordering of $B$ and $C$ depends on how linear $B$ is relative to how noisy $A$ is, so neither $M_\text{D,B}$ nor $M_\text{D,C}$ alone is sufficient. Thus, while ordering remains based on ascending $M_\text{D}$, it must be applied adaptively, updating the effective $M_\text{D}$ after each stage is added.

The utility of $M_\text{D}$ relative to $M_\text{N}$ and $M_\text{L}$ depends on the system-level objective. When noise figure and linearity are independently specified---for example, when a receiver must meet both a sensitivity requirement and an IIP3 requirement---$M_\text{N}$ and $M_\text{L}$ are the natural design handles. When DR is the single binding constraint, $M_\text{D}$ is the appropriate design measure, since it captures the joint contribution of noise and linearity to DR in one number. In particular, $M_\text{D}$ resolves trade-offs that $M_\text{N}$ and
$M_\text{L}$ cannot settle individually: if Device~\#1 is quieter but less linear than Device~\#2, neither $M_\text{N}$ nor $M_\text{L}$ alone determines which degrades cascade DR less, whereas the device with the smaller $M_\text{D}$ should unambiguously be placed first. The application of $M_\text{N}$, $M_\text{L}$, and $M_\text{D}$ to cascade ordering is illustrated in Sidebar~C.

\section{\protect\label{sec:System-Efficiency}Power Efficiency, Waste Factor, and Efficiency and Waste Measures}

Power-efficiency metrics are widely used in RF system design. In many practical implementations, these quantities vary only weakly with instantaneous signal level: transmitters typically operate near a specified output power, and receivers employ automatic gain control to maintain consistent signal levels at the ADC. As a result, efficiency and waste factor are often treated as average or near-constant device characteristics and used accordingly in system modeling and link-budget analysis \cite{Rappaport_WasteFactor_2024,Rappaport2025,10436843,Rappaport_Globecom2024,Rappaport_2014,Rappaport_MJ2024}. In this work, however, we focus on the instantaneous behavior of these metrics, where both $\eta$ and $W$ depend directly on the instantaneous input and output power. This signal-dependent perspective is essential when connecting power efficiency to dynamic range, EVM, and the system-aware measures developed in earlier sections.

The efficiency with which an RF device converts dc power into useful output power is commonly quantified by $\eta=P_{\text{out}}/P_{\text{dc}}$, and the related power-added efficiency (PAE), $\eta_\text{PAE}=(P_{\text{out}}-P_{\text{in}})/P_{\text{dc}}= \eta(G-1)/G$, where $P_\text{dc}$ is the dc power consumption and $G$ is the gain. A complementary perspective is provided by the waste factor $W$, defined in ~\cite[equation (3)]{Rappaport_WasteFactor_2024} as
\begin{equation}
W=\frac{P_{\text{non-signal}}+P_{\text{out}}}{P_{\text{out}}}
=\frac{1}{\eta}+\frac{1}{G},
\label{eq:waste-factor}
\end{equation}
where \(P_{\text{non-signal}} = (P_\text{dc}+P_\text{in}) - P_\text{out}\) represents the difference between the total supplied power and the useful output signal power. Whereas $\eta$ captures the fraction of dc power converted to useful output, $W$ quantifies the total supplied power required to deliver a unit of output signal power. The two metrics are related, and each provides insight depending on whether the designer wishes to emphasize efficiency or inefficiency.
For cascaded devices, overall efficiency and waste factor follow directly from definitions
\begin{equation}
\left\{
\begin{aligned}
\eta_{12}
&= \frac{P_{\text{out},2}}{P_{\text{dc},1}+P_{\text{dc},2}} \\
W_{12}
&= \frac{P_{\text{dc},1}+P_{\text{dc},2}+P_{\text{in},1}}{P_{\text{out},2}}
= \frac{1}{\eta_{12}}+\frac{1}{G_1G_2}.
\end{aligned}
\right.
\label{eq:cascade_eta}
\end{equation}

Because these expressions are independent of device ordering, neither $\eta$ nor $W$ can guide architectural choices in the way noise factor, IP, or dynamic range can. This invariance reveals a key limitation: evaluated at an arbitrary operating point, $\eta$ and $W$ cannot serve as system-aware measures and thus cannot determine the optimal cascade order or prevent nonphysical or trivial design outcomes.

To obtain system-aware quantities—analogous to the noise and linearity measures defined previously—we evaluate $\eta$ and $W$ at the maximum allowable signal level, determined by linearity constraints such as IP or OP. This yields the extremal values $\eta_\text{max}=P_\text{out}^\text{max}/P_\text{dc}$ and $W_\text{min}=1/\eta_\text{max}+1/G$. For cascaded devices, the corresponding expressions become
\begin{equation}
\left\{
\begin{aligned}
\frac{1}{\eta^\text{max}_\text{12}} &
= \frac{1}{G_2\eta^{max}_{1}} \left(1+\frac{P_{\text{dc},2}}{P_{\text{dc},1}}\right)
+ \frac{1}{\eta^\text{max}_{2}}
\left(1+\frac{P_{\text{dc},1}}{P_{\text{dc},2}}\right) \\
W_\text{12}^\text{min} &=\frac{W_\text{1}^\text{min}G_\text{1}-1}{G_\text{2}G_\text{1}}\left(1+\frac{P_\text{dc,2}}{P_\text{dc,1}}\right)\\
 & +\frac{W_\text{2}^\text{min}G_\text{2}-1}{G_\text{2}}\left(1+\frac{P_\text{dc,1}}{P_\text{dc,2}}\right)+\frac{1}{G_\text{1}G_\text{2}}.
\end{aligned}
\right.
\label{eq:cascade-eta-max}
\end{equation}

These extremal forms allow the construction of the \textit{efficiency measure} and the reciprocal \textit{waste measure},
\begin{equation}
\left\{
\begin{aligned}
M_\eta &\overset{\triangle}{=}
\eta^\text{max}\left(1-G^{-1}\right)P_\text{dc}\\
M_W &=  
M_\eta^{-1}.
\end{aligned}
\right.
\label{eq:eta-measure}
\end{equation}

These measures depend on the maximum usable input or output power rather than a nominal operating point, making them directly comparable across devices with different gains, biases, or linearity limits. As with $M_\text{N}$ and $M_\text{L}$, ordering devices by increasing $M_\eta$ (or equivalently, decreasing $M_W$) maximizes system-level performance when $G>1$.
Viewed in this way, $M_\eta$ and $M_W$ play for power-efficiency optimization the same role that $M_N$ and $M_L$ play for noise and linearity: they transform familiar metrics into quantities that meaningfully describe performance in cascaded, system-constrained scenarios. Applications of $M_\eta$ and $M_W$ are illustrated in Sidebar~D.

\section{\protect\label{sec:System-Factor}Operating Efficiency}

Section~\ref{sec:System-Efficiency} showed that device power efficiency—quantified by metrics such as $\eta$, $\eta_{\text{PAE}}$, and $W$—depends on the input signal power level. In practical RF systems, input power varies continuously: transmitter power fluctuates with modulation, while receiver input power changes with mobility, interference, and channel conditions. The allowable range of input signal power, denoted DRo in Fig.~\ref{fig:EVM}, is bounded by maximum tolerable EVM, which is set by the target data rate and BER via \eqref{eq:BER-defined} and \eqref{eq:Data-rate-F-DR-IP-1}. Within this range, EVM degrades at low powers due to noise and at high powers due to nonlinear distortion. Improving EVM by reducing noise, enhancing linearity, suppressing phase noise, or increasing transmit power generally increases power consumption and waste. Conversely, optimizing for power efficiency alone may degrade signal quality, since conventional efficiency metrics do not directly reflect how effectively information is conveyed. As a result, signal-quality metrics (BER, EVM, DRo) and power-efficiency metrics ($\eta$, $\eta_{\text{PAE}}$, $W$) are not inherently aligned.

This disconnect is amplified by the statistical nature of signal power. In a receiver, DRo typically spans several decades of input power, and instantaneous efficiency may be near zero at low powers and peak near the upper end of the range. If low input powers occur frequently, the receiver will most often operate inefficiently; if higher powers are more likely, average efficiency improves. In contrast, transmitter output power is usually concentrated near its mean value, with infrequent peaks approaching the maximum. These observations emphasize that signal power is not fixed but distributed across DRo according to a probability density function (PDF), motivating efficiency metrics that account for both power range and likelihood of occurrence.

To address this, we introduce \emph{operating efficiency}, $\eta_\text{o}$, which captures both the variation of instantaneous power efficiency and the statistical distribution of signal power. Operating efficiency is defined as the probability-weighted average of instantaneous efficiency over the allowable input power range
\begin{equation}
\eta_\text{o} \triangleq \int_{P_\text{in}^\text{min}}^{P_\text{in}^\text{max}} \eta(P_\text{in})\, \mathscr{P}(P_\text{in}) \mathrm{d}P_\text{in},
\label{eq:SF-defined}
\end{equation}
where $\eta(P_\text{in}) = G(P_\text{in})P_\text{in}/P_\text{dc}(P_\text{in})$ and $\mathscr{P}(P_\text{in})$ is the PDF of the input signal power. When $\mathscr{P}(P_\text{in})$ integrates to unity over the interval, $\eta_\text{o}$ corresponds to the expected value of $\eta(P_\text{in})$.

The integration limits encode $\text{DRo}=P_\text{in}^\text{max}/P_\text{in}^\text{min}$ in \eqref{eq:SF-defined} and through a variable substitution, e.g., $u=P_\text{in}/P_\text{in}^\text{min}$ or $v=P_\text{in}/P_\text{in}^\text{max}$, can be rearranged to show how DRo impacts the integration limits explicitly. Because DRo is linked to EVM and data rate through $\eta_{\text{DR}} \approx \text{EVM}^3$ and \eqref{eq:Data-rate-F-DR-IP-1}, operating efficiency is inherently connected to both signal quality and throughput.

The definition of $\eta_\text{o}$ is not restricted to fixed biasing or fixed modulation. Any dependence of $P_\text{dc}$ on $P_\text{in}$ is captured through $\eta(P_\text{in})$, and the statistical formulation naturally accommodates adaptive modulation and variable-rate systems. As such, $\eta_\text{o}$ provides a flexible framework for evaluating aggregate efficiency under realistic operating conditions.

The concept of averaging instantaneous efficiency over a signal's power distribution has prior precedents in PA design. Raab~\cite{Raab1986,Raab1987} computed mean efficiency for Class~G and Doherty PAs by folding the analytical $\eta(P_\text{in})$ curve against a Rayleigh envelope distribution, and Cripps~\cite{Cripps2006} extended this to additional topologies. Operating efficiency $\eta_\text{o}$ generalises these results in three respects: it accepts any $\eta(P_\text{in})$ regardless of PA topology; the integration limits are set by the noise floor and linearity ceiling of the operating dynamic range, explicitly coupling efficiency to the measures developed earlier in this paper; and the signal distribution $\mathscr{P}(P_\text{in})$ is left arbitrary, accommodating adaptive modulation and variable-rate systems. These extensions allow $\eta_\text{o}$ to serve as a unified metric across device types, signal environments, and system configurations beyond the scope of PA-focused efficiency analyses.

Following the methodology used to define system-aware measures in earlier sections, we evaluate the operating efficiencies
\begin{equation}
\left\{
\begin{aligned}
\eta_\text{o,12} &= \int_{P_\text{in,12}^\text{min}}^{P_\text{in,12}^\text{max}} \eta_{12}(P_\text{in})\,\mathscr{P}(P_\text{in})\,\mathrm{d}P_\text{in} \\
\eta_\text{o,21} &= \int_{P_\text{in,21}^\text{min}}^{P_\text{in,21}^\text{max}} \eta_{21}(P_\text{in})\,\mathscr{P}(P_\text{in})\,\mathrm{d}P_\text{in}
\end{aligned}
\right.
\label{eq:cascaded-OE_1}
\end{equation}
for the two cascade orderings in Fig.~\ref{fig:Two-possible-cascaded}. Since $\eta_{12}(P_\text{in})=\eta_{21}(P_\text{in})$, any difference between $\eta_\text{o,12}$ and $\eta_\text{o,21}$ arises solely from differences in the integration limits. When the actual signal-power distribution lies entirely within both ranges, $\eta_\text{o,12}=\eta_\text{o,21}$, as is typically the case in transmitter chains. In receiver systems, however, interference or weak signals may cause partial overlap, leading to different operating efficiencies.

To illustrate when $\eta_{\text{o},12} \neq \eta_{\text{o},21}$, consider two receiver cascades in Fig.~\ref{fig:Two-possible-cascaded}, where $F_{12} < F_{21}$ while the linearity ceiling is the same for both orderings. If the input signal PDF has probability mass in the interval $[kTBF_{12},\, kTBF_{21}]$, which corresponds to weak signals below the sensitivity limit of Cascade~2$\to$1, then $\eta_\text{o,12}$ includes integration over this region and captures the associated efficiency contribution, while $\eta_\text{o,21}$ does not. This yields $\eta_{\text{o,12}} > \eta_{\text{o,21}}$. Conversely, if the linearity ceilings of the two cascades differ, and the signal PDF has mass above the lower of the two ceilings, then also $\eta_{\text{o},12} \neq \eta_{\text{o},21}$. The difference between $\eta_\text{o,12}$ and $\eta_\text{o,21}$ vanishes when the signal PDF has no mass outside the intersection of the two DRo ranges, which is why transmitter chains---where the signal power is well-controlled and concentrated near its mean---typically show $\eta_{\text{o},12} = \eta_{\text{o},21}$.

Worked examples of $\eta_\text{o}$ for common device types, including a linear LNA and a Doherty PA under uniform and M-QAM signaling, are provided in Sidebar~E. As shown in Sidebar~E, the choice of input power distribution $\mathscr{P}(P_\text{in})$ directly determines $\eta_\text{o}$, and $\mathscr{P}(P_\text{in})$ impact depends on the device type. For a linear amplifier, \eqref{eq:SF-M-QAM-lin-amp} shows that $\eta_\text{o} = G \times P_\text{in}^\text{ave}/P_\text{dc}$, so $\eta_\text{o}$ depends only on the mean input power, not on the distribution shape. The uniform distribution assumed in Sidebar~E-A spans the full DRo including power levels near the noise floor, which lowers $P_\text{in}^\text{ave}$ and therefore $\eta_\text{o}$, as reflected in the 2\% quoted for the LNA example. For nonlinear devices such as Doherty PAs, the full distribution shape matters because $\eta(P_\text{in})$ is non-monotone. The sensitivity to modulation order is visible in Sidebar~E-D: $\eta_\text{o}$ decreases from 29\% for 64-QAM to 26\% for 256-QAM because higher-order constellations have a larger peak-to-average power ratio, shifting more probability mass toward lower-power symbols where the PA operates less efficiently.

The results in Sidebar~E also show that LNAs exhibit significantly lower $\eta_\text{o}$ than PAs, but this should not motivate conclusions about component quality. A fair comparison of $\eta_\text{o}$ across different devices requires the same $\mathscr{P}(P_\text{in})$ to be used for all devices. Furthermore, a device operating over a larger DRo generally exhibits lower $\eta_\text{o}$ because its efficiency is averaged over a wider power range that includes regions of very low instantaneous efficiency. A low $\eta_\text{o}$ may therefore be a direct consequence of a system design requiring a large dynamic range rather than an indication of a poorly designed component.

In summary, operating efficiency $\eta_\text{o}$ depends jointly on device behavior, allowable signal-power range, and the statistics of $P_\text{in}$. It therefore cannot be reduced to a device-only measure. Achieving high $\eta_\text{o}$ requires shaping the dynamic range—through noise, linearity, and dynamic-range metrics—so that regions of high instantaneous efficiency align with regions of high signal probability. 

\section{\protect\label{sec:Conclusions}Conclusion}

This article revisited the role of performance metrics and measures in RF communication systems, emphasizing the distinction between device-level metrics and system-aware measures that reflect cascading behavior and physical trade-offs. By reexamining classical quantities such as noise factor, linearity, dynamic range, and power efficiency within a unified system context, the paper highlighted how familiar metrics can yield deeper insight when interpreted through system-level measures rather than in isolation. Table~\ref{tab:summary} summarizes all metrics and measures discussed in this article.
\begin{table}[]
\centering
\caption{Summary of metrics and measures discussed in this article. ``Metrics'' describe device-level performance in isolation; ``measures'' combine multiple device metrics to reflect their joint impact on system performance. The ``Use when'' column gives the primary application context.}
\label{tab:summary}
\renewcommand{\arraystretch}{1.3}
\begin{tabular}{@{}p{0.8cm}@{\hspace{5pt}}|@{\hspace{3pt}}p{1cm}@{\hspace{3pt}}|@{\hspace{3pt}}p{2.2cm}@{\hspace{3pt}}|@{\hspace{3pt}}p{3.9cm}@{\hspace{3pt}}}
\hline
\rowcolor[HTML]{31668D}\textcolor{white}{\textbf{Symbol}}
& \textcolor{white}{\textbf{Type}}
& \textcolor{white}{\textbf{Name}}
& \textcolor{white}{\textbf{Use when}} \\
\hline\hline
\rowcolor[HTML]{D6E8F5}$F$ & Metric & Noise factor
& Characterizing SNR degradation due to device noise \\
\rowcolor[HTML]{D6E8F5}IP, OP & Metric & Intercept and compression points
& Characterizing single-device linearity (IP$_{1\text{dB}}$,
IIP3, OP$_{1\text{dB}}$, OIP3) \\
\rowcolor[HTML]{D6E8F5}DR & Metric & Dynamic range
& Characterizing single-device operating window \\
\rowcolor[HTML]{D6E8F5}$\eta$ & Metric & Power efficiency
& Power efficiency at a fixed input power\\
\rowcolor[HTML]{D6E8F5}$W$ & Metric & Waste factor
& Power waste at a fixed input power \\
\hline
\rowcolor[HTML]{EAF4D3}$M_\text{N}$ & Measure & Noise measure
& Ascending order minimizes cascade $F$ \\
\rowcolor[HTML]{EAF4D3}$M_\text{L}$ & Measure & Linearity measure
& Ascending order maximizes cascade IP \\
\rowcolor[HTML]{EAF4D3}$M_\text{D}$ & Measure & Dynamic-range measure
& Ascending order maximizes cascade DR \\
\rowcolor[HTML]{EAF4D3}$M_\eta$ & Measure & Efficiency measure
& Ascending order maximizes cascade $\eta$ \\
\rowcolor[HTML]{EAF4D3}$M_W$ & Measure & Waste measure
& Descending order minimizes cascade $W$ \\
\rowcolor[HTML]{EAF4D3}$\eta_\text{o}$ & Measure & Operating efficiency
& Evaluating aggregate efficiency under realistic signal statistics and comparing cascade orderings. \\
\hline
\end{tabular}
\end{table}

Within this framework, operating efficiency was introduced as a system-level descriptor that links power efficiency, dynamic range, and achievable data rate under practical operating conditions. Operating efficiency is intended to complement existing metrics by clarifying how signal statistics, modulation, and implementation constraints jointly influence power consumption in RF systems.

Although the examples in this paper focus on narrowband modulated signals, the proposed framework is not inherently restricted to narrowband operation. The definitions of efficiency and waste measures depend on maximum usable signal power and signal statistics rather than bandwidth per se. In wideband or impulse-based signaling schemes, power efficiency improvements arise from reduced average signal power and relaxed linearity requirements. In such cases, the quantities \(P_\text{in}^\text{max}\) and \(P_\text{out}^\text{max}\) must be defined with respect to peak constraints, spectral masks, and receiver noise bandwidth, rather than steady-state sinusoidal metrics. While a detailed treatment of these effects is beyond the scope of this work, the same system-aware measures, including operating efficiency, can be applied once appropriate signal-dependent limits are established.

\section*{Sidebar A: Approaches to Reducing $\eta_\text{DR}$}
Since $\eta_\text{DR}=\text{DRo}/\text{DR}_\text{max}$ represents the fraction of the total available dynamic range that is usable when EVM is constrained to support a target spectral efficiency $R/B$, the increase in data rate, which reduces  $\eta_\text{DR}$, can be achieved in two ways:

\textbf{1. Reducing $\text{DRo}$:} Increasing the required SNR (raising $P_\text{in}^\text{min}$) and tightening linearity constraints (lowering $P_\text{in}^\text{max}$) reduces DRo and, consequently, $\eta_\text{DR}$. Rewriting \eqref{eq:Data-rate-F-DR-IP-1} in terms of SNR and power backoff $P_\text{bo}^\text{IM}$ gives
\begin{align}
\frac{P_\text{bo}^\text{IM}}{\text{SNR}} & =\Gamma^{-3/2}\left(2^{R/B}-1\right)^{-3/2}.
\end{align}
While this approach leaves the circuits unchanged, it reduces communication range and tolerance to strong interference.

\textbf{2. Increasing $\text{DR}_\text{max}$:} Alternatively, improving circuit noise performance and linearity (i.e., increasing SNR and IIP3) raises $\text{DR}_\text{max}$, albeit at the cost of higher power consumption and design complexity.

Compared to reducing DRo, increasing $\text{DR}_\text{max}$ provides a more sustainable path for long-term data-rate scaling. Part of the associated power cost can be mitigated through antenna gain and array techniques \cite{Bjornson2017,Rappaport2025,Rappaport_2011,Rappaport2012,BelostotskiAhmadi_2017,BelostotskiRadpour2024,BelostotskiXie2025} at the expense of increased system complexity.

\section*{Sidebar B: Noise- and Linearity-Limited Cascaded Dynamic Range}

If the cascaded DRo is limited solely by noise (i.e., $P_\text{in}^\text{max}$ lies below the cascade linearity limit), then
\begin{equation}
\text{DRo}_\text{12}^{-1}=\text{DRo}_\text{1}^{-1}+\left (1-F_\text{2}^{-1}\right)\text{DRo}_\text{2}^{-1}.\label{eq:Dr-noise-limited}
\end{equation}
Conversely, if the DRo is limited by linearity (i.e., $P_\text{in}^\text{min}$ lies well above the noise floor), the cascaded DRo reduces to
\begin{equation}
\text{DRo}_\text{12}^{-1}=\text{DRo}_\text{1}^{-1}+\text{DRo}_\text{2}^{-1}.\label{eq:DR-linearity-limited}
\end{equation}

\section*{Sidebar C: Cascade Ordering Using Noise, Linearity, and Dynamic-Range Measures}

To illustrate how $M_\text{N}$, $M_\text{L}$, and $M_\text{D}$ can be used for cascade ordering, consider the three amplifiers in Table~\ref{tab:Sidebar-C}. These amplifiers may represent finalized designs or intermediate candidates in a design or optimization flow requiring system-level evaluation. 
\setlength{\tabcolsep}{3pt}
\renewcommand{\arraystretch}{1.2}
\begin{table*}
\centering
\caption{\protect\label{tab:Sidebar-C}Device parameters and measures for the cascade ordering example. Cell shading encodes per-column value via the Viridis colormap (dark purple = worst value, bright yellow = best value), where best means lowest for $F$ and $M_\text{N}$, and highest for $G$, $\text{IIP3}$, $M_\text{L}$, and $M_\text{D}$.}
\begin{tabular}{c||c|c|c||c|c|c}
\hline
{\footnotesize Device}
& {\footnotesize $F$ (dB)}
& {\footnotesize $G$ (dB)}
& {\footnotesize $\text{IIP3}$ (dBm)}
& {\footnotesize $M_\text{N}$}
& {\footnotesize $M_\text{L}$ (mW)}
& {\footnotesize $M_\text{D}$ (mW)}
\tabularnewline
\hline\hline
{\footnotesize \textcolor{ampone}{Amp.~\#1}}
& {\footnotesize \cellcolor[HTML]{D2E21B}\textcolor{black}{3.00}}
& {\footnotesize \cellcolor[HTML]{FDE725}\textcolor{black}{18.0}}
& {\footnotesize \cellcolor[HTML]{440154}\textcolor{white}{$-$15.0}}
& {\footnotesize \cellcolor[HTML]{FDE725}\textcolor{black}{1.01}}
& {\footnotesize \cellcolor[HTML]{FDE725}\textcolor{black}{1.96}}
& {\footnotesize \cellcolor[HTML]{423F85}\textcolor{white}{3.95}}
\tabularnewline
\hline
{\footnotesize \textcolor{amptwo}{Amp.~\#2}}
& {\footnotesize \cellcolor[HTML]{440154}\textcolor{white}{10.0}}
& {\footnotesize \cellcolor[HTML]{365C8D}\textcolor{white}{8.00}}
& {\footnotesize \cellcolor[HTML]{FDE725}\textcolor{black}{$-$5.00}}
& {\footnotesize \cellcolor[HTML]{440154}\textcolor{white}{10.7}}
& {\footnotesize \cellcolor[HTML]{98D83E}\textcolor{black}{1.68}}
& {\footnotesize \cellcolor[HTML]{FDE725}\textcolor{black}{19.6}}
\tabularnewline
\hline
{\footnotesize \textcolor{ampthree}{Amp.~\#3}}
& {\footnotesize \cellcolor[HTML]{FDE725}\textcolor{black}{2.50}}
& {\footnotesize \cellcolor[HTML]{440154}\textcolor{white}{4.00}}
& {\footnotesize \cellcolor[HTML]{21918C}\textcolor{white}{$-$10.0}}
& {\footnotesize \cellcolor[HTML]{ECE51B}\textcolor{black}{1.29}}
& {\footnotesize \cellcolor[HTML]{440154}\textcolor{white}{0.151}}
& {\footnotesize \cellcolor[HTML]{440154}\textcolor{white}{0.347}}
\tabularnewline
\hline
\end{tabular}
\end{table*}
Table~\ref{tab:Sidebar-C-cascade} shows all pairwise cascades and their resultant cascaded $F$ ($F_\text{casc}$), IIP3 ($\text{IIP3}_\text{casc}$), and $\text{DR}_\text{max}$ ($\text{DR}_\text{max,casc}$). As expected, $M_\text{N}$-, $M_\text{L}$-, and $M_\text{D}$-based ordering improves noise, linearity, and dynamic range, respectively. In the pairwise comparisons, $M_\text{D}$-based ordering maximizes $\text{DR}_\text{max,casc}$ in all three cases where it is applied. The table also shows the cascaded measures ($M_\text{N,casc}$, $M_\text{L,casc}$, and $M_\text{D,casc}$), which can be computed directly from their definitions or from the cascade formulas indicated below the table.
\setlength{\tabcolsep}{3pt}
\renewcommand{\arraystretch}{1.2}
\begin{table*}
\centering
\caption{\protect\label{tab:Sidebar-C-cascade}Pairwise cascade metrics. Cell shading encodes per-column value as in Table~\ref{tab:Sidebar-C}.}
\begin{tabular}{c||c|c|c|c||c|c|c||c}
\hline
{\footnotesize Pair}
& {\footnotesize Ordering}
& {\footnotesize $F_\text{casc}$ (dB)}
& {\footnotesize $\text{IIP3}_\text{casc}$ (dBm)}
& {\footnotesize $\text{DR}_\text{max,casc}$ (dBHz)}
& {\footnotesize $M_\text{N,casc}$}
& {\footnotesize $M_\text{L,casc}$ (mW)}
& {\footnotesize $M_\text{D,casc}$ (mW)}
& {\footnotesize Ordered by}
\tabularnewline
\hline\hline
\multirow{2}{*}{{\footnotesize \textcolor{ampone}{\#1}
  vs \textcolor{amptwo}{\#2}}}
& {\footnotesize \textcolor{ampone}{\#1}$\to$\textcolor{amptwo}{\#2}}
& {\footnotesize \cellcolor[HTML]{E7E419}\textcolor{black}{3.30}}
& {\footnotesize \cellcolor[HTML]{355E8D}\textcolor{white}{$-$23.6}}
& {\footnotesize \cellcolor[HTML]{24878E}\textcolor{white}{147}}
& {\footnotesize \cellcolor[HTML]{F6E620}\textcolor{black}{1.14}}
& {\footnotesize \cellcolor[HTML]{A2DA37}\textcolor{black}{1.72}}
& {\footnotesize \cellcolor[HTML]{443983}\textcolor{white}{3.68}}
& {\footnotesize $M_\text{N}$, $M_\text{D}$}
\tabularnewline
\cline{2-9}
& {\footnotesize \textcolor{amptwo}{\#2}$\to$\textcolor{ampone}{\#1}}
& {\footnotesize \cellcolor[HTML]{440154}\textcolor{white}{10.1}}
& {\footnotesize \cellcolor[HTML]{31668E}\textcolor{white}{$-$23.1}}
& {\footnotesize \cellcolor[HTML]{440154}\textcolor{white}{141}}
& {\footnotesize \cellcolor[HTML]{48186A}\textcolor{white}{9.18}}
& {\footnotesize \cellcolor[HTML]{FDE725}\textcolor{black}{1.96}}
& {\footnotesize \cellcolor[HTML]{FDE725}\textcolor{black}{19.9}}
& {\footnotesize $M_\text{L}$}
\tabularnewline
\hline
\multirow{2}{*}{{\footnotesize \textcolor{ampone}{\#1}
  vs \textcolor{ampthree}{\#3}}}
& {\footnotesize \textcolor{ampone}{\#1}$\to$\textcolor{ampthree}{\#3}}
& {\footnotesize \cellcolor[HTML]{FDE725}\textcolor{black}{3.03}}
& {\footnotesize \cellcolor[HTML]{440154}\textcolor{white}{$-$28.2}}
& {\footnotesize \cellcolor[HTML]{453581}\textcolor{white}{143}}
& {\footnotesize \cellcolor[HTML]{FDE725}\textcolor{black}{1.01}}
& {\footnotesize \cellcolor[HTML]{440256}\textcolor{white}{0.237}}
& {\footnotesize \cellcolor[HTML]{440154}\textcolor{white}{0.479}}
& {\footnotesize $M_\text{N}$}
\tabularnewline
\cline{2-9}
& {\footnotesize \textcolor{ampthree}{\#3}$\to$\textcolor{ampone}{\#1}}
& {\footnotesize \cellcolor[HTML]{DFE318}\textcolor{black}{3.37}}
& {\footnotesize \cellcolor[HTML]{1E9D89}\textcolor{white}{$-$19.5}}
& {\footnotesize \cellcolor[HTML]{67CC5C}\textcolor{black}{151}}
& {\footnotesize \cellcolor[HTML]{F4E61E}\textcolor{black}{1.18}}
& {\footnotesize \cellcolor[HTML]{B2DD2D}\textcolor{black}{1.76}}
& {\footnotesize \cellcolor[HTML]{443B84}\textcolor{white}{3.84}}
& {\footnotesize $M_\text{L}$, $M_\text{D}$}
\tabularnewline
\hline
\multirow{2}{*}{{\footnotesize \textcolor{amptwo}{\#2}
  vs \textcolor{ampthree}{\#3}}}
& {\footnotesize \textcolor{amptwo}{\#2}$\to$\textcolor{ampthree}{\#3}}
& {\footnotesize \cellcolor[HTML]{440154}\textcolor{white}{10.1}}
& {\footnotesize \cellcolor[HTML]{2CB17E}\textcolor{black}{$-$18.2}}
& {\footnotesize \cellcolor[HTML]{2B748E}\textcolor{white}{146}}
& {\footnotesize \cellcolor[HTML]{440154}\textcolor{white}{9.74}}
& {\footnotesize \cellcolor[HTML]{440154}\textcolor{white}{0.224}}
& {\footnotesize \cellcolor[HTML]{482475}\textcolor{white}{2.41}}
& {\footnotesize ---}
\tabularnewline
\cline{2-9}
& {\footnotesize \textcolor{ampthree}{\#3}$\to$\textcolor{amptwo}{\#2}}
& {\footnotesize \cellcolor[HTML]{2A778E}\textcolor{white}{7.29}}
& {\footnotesize \cellcolor[HTML]{FDE725}\textcolor{black}{$-$12.5}}
& {\footnotesize \cellcolor[HTML]{FDE725}\textcolor{black}{154}}
& {\footnotesize \cellcolor[HTML]{20A486}\textcolor{white}{4.65}}
& {\footnotesize \cellcolor[HTML]{2F6C8E}\textcolor{white}{0.828}}
& {\footnotesize \cellcolor[HTML]{3F4889}\textcolor{white}{4.68}}
& {\footnotesize $M_\text{N}$, $M_\text{L}$, $M_\text{D}$}
\tabularnewline
\hline
\end{tabular}
\noindent\parbox{\linewidth}{\centering\footnotesize\vspace{4pt}
Cascaded measures: $M_{\text{N},ij}=M_{\text{N},i}+\left(M_{\text{N},j}-M_{\text{N},i}\right)\dfrac{G_j-1}{G_iG_j-1}$; 
$M_{\text{L},ij}^{-1}=M_{\text{L},j}^{-1}+\left(M_{\text{L},i}^{-1}-M_{\text{L},j}^{-1}\right)\dfrac{G_i-1}{G_iG_j-1}$; $M_{\text{D},ij}\approx(M_{\text{N},ij}+1)M_{\text{L},ij}$.
$\text{DR}_\text{max}\text{ (dBHz)}=[\text{IIP3}_\text{casc} - F_\text{casc} - (kT)]_\text{dB}$ with $(kT)_\text{dB}=-174\,\text{dBm/Hz}$}
\end{table*}

The need for an adaptive procedure for three or more devices identified in Subsection~\ref{subsec:Dynamic-Range} follows from the structure of the cascade formulas. Whereas ascending $M_\text{N}$ and $M_\text{L}$ ordering yield noise- and linearity-optimal orderings, respectively, Subsection~\ref{subsec:Dynamic-Range} shows that the preferred $M_\text{D}$-based ordering of adjacent devices depends on the cumulative  $F$ ($F_\text{prec}$) and $G$ ($G_\text{prec}$) of preceding stages. This dependence can be understood as follows: a large value of $G_{\text{prec}}F_{\text{prec}}$ increases the relative weighting of linearity, favouring the sub-ordering with higher $\text{IIP3}_\text{casc}$, while a small $G_{\text{prec}}F_{\text{prec}}$ shifts the balance toward noise, favouring lower $F_\text{casc}$.

For ordering amplifiers in Table~\ref{tab:Sidebar-C}, ascending $M_\text{D}$ ordering places Amp.~\#3 first. To continue with ascending ordering, we need to find a second device such that total $M_\text{D}$ of first two devices is lower than that of the third device. Therefore, we use the cascaded $M_\text{D}$ from Table~\ref{tab:Sidebar-C-cascade} for each candidate second device (\#1 or \#2) to find $M_\text{D,31}=3.84\,\text{mW}< M_\text{D,2}=19.6\,\text{mW}$ and $M_\text{D,32}=4.68\,\text{mW} \not< M_\text{D,1}=3.95\,\text{mW}$. Only the first inequality is satisfied, confirming Amp.~\#1 as the second device and Amp.~\#2 last, yielding the optimum \#3$\to$\#1$\to$\#2 order. Table~\ref{tab:Sidebar-C-3way} confirms this is optimal, with the $M_\text{N}$-, $M_\text{L}$-based, and other orderings achieving lower $\text{DR}_\text{max}$, as shown in the table.
\setlength{\tabcolsep}{3pt}
\renewcommand{\arraystretch}{1.2}
\begin{table*}
\centering
\caption{\protect\label{tab:Sidebar-C-3way}All six three-device orderings ranked by $\text{DR}_\text{max}$ (dBHz). Cell shading encodes per-column value as in Table~\ref{tab:Sidebar-C}.}
\begin{tabular}{c||c|c|c||c}
\hline
{\footnotesize Ordering}
& {\footnotesize $F_\text{casc}$ (dB)}
& {\footnotesize $\text{IIP3}_\text{casc}$ (dBm)}
& {\footnotesize $\text{DR}_\text{max}$ (dBHz)}
& {\footnotesize Note}
\tabularnewline
\hline\hline
{\footnotesize %
  \textcolor{ampthree}{\#3}$\to$%
  \textcolor{ampone}{\#1}$\to$%
  \textcolor{amptwo}{\#2}}
& {\footnotesize \cellcolor[HTML]{DFE318}\textcolor{black}{3.49}}
& {\footnotesize \cellcolor[HTML]{DAE319}\textcolor{black}{$-$27.7}}
& {\footnotesize \cellcolor[HTML]{FDE725}\textcolor{black}{143}}
& {\footnotesize $M_\text{D}$ order}
\tabularnewline
\hline
{\footnotesize %
  \textcolor{ampone}{\#1}$\to$%
  \textcolor{ampthree}{\#3}$\to$%
  \textcolor{amptwo}{\#2}}
& {\footnotesize \cellcolor[HTML]{FDE725}\textcolor{black}{3.15}}
& {\footnotesize \cellcolor[HTML]{25AB82}\textcolor{white}{$-$30.7}}
& {\footnotesize \cellcolor[HTML]{7AD151}\textcolor{black}{140}}
& {\footnotesize $M_\text{N}$ order}
\tabularnewline
\hline
{\footnotesize %
  \textcolor{ampthree}{\#3}$\to$%
  \textcolor{amptwo}{\#2}$\to$%
  \textcolor{ampone}{\#1}}
& {\footnotesize \cellcolor[HTML]{2A778E}\textcolor{white}{7.34}}
& {\footnotesize \cellcolor[HTML]{FDE725}\textcolor{black}{$-$27.2}}
& {\footnotesize \cellcolor[HTML]{54C568}\textcolor{black}{139}}
& {\footnotesize $M_\text{L}$ order}
\tabularnewline
\hline
{\footnotesize %
  \textcolor{amptwo}{\#2}$\to$%
  \textcolor{ampthree}{\#3}$\to$%
  \textcolor{ampone}{\#1}}
& {\footnotesize \cellcolor[HTML]{440154}\textcolor{white}{10.1}}
& {\footnotesize \cellcolor[HTML]{EAE51A}\textcolor{black}{$-$27.5}}
& {\footnotesize \cellcolor[HTML]{1F988B}\textcolor{white}{136}}
& {\footnotesize ---}
\tabularnewline
\hline
{\footnotesize %
  \textcolor{ampone}{\#1}$\to$%
  \textcolor{amptwo}{\#2}$\to$%
  \textcolor{ampthree}{\#3}}
& {\footnotesize \cellcolor[HTML]{F1E51D}\textcolor{black}{3.30}}
& {\footnotesize \cellcolor[HTML]{440154}\textcolor{white}{$-$36.2}}
& {\footnotesize \cellcolor[HTML]{2A788E}\textcolor{white}{134}}
& {\footnotesize ---}
\tabularnewline
\hline
{\footnotesize %
  \textcolor{amptwo}{\#2}$\to$%
  \textcolor{ampone}{\#1}$\to$%
  \textcolor{ampthree}{\#3}}
& {\footnotesize \cellcolor[HTML]{440154}\textcolor{white}{10.1}}
& {\footnotesize \cellcolor[HTML]{440154}\textcolor{white}{$-$36.2}}
& {\footnotesize \cellcolor[HTML]{440154}\textcolor{white}{128}}
& {\footnotesize ---}
\tabularnewline
\hline
\end{tabular}
\noindent\parbox{\linewidth}{\centering\footnotesize\vspace{4pt}
$\text{DR}_\text{max}\text{ (dBHz)}=[\text{IIP3}_\text{casc} - F_\text{casc} - (kT)]_\text{dB}$
with $(kT)_\text{dB}=-174\,\text{dBm/Hz}$}
\end{table*}

\section*{Sidebar D: Applications of Efficiency and Waste Measures}

To illustrate why efficiency and waste measures are needed, consider the two amplifiers in Table~\ref{tab:Example-of-Two-Amps}.
\setlength{\tabcolsep}{3pt}
\renewcommand{\arraystretch}{1.2}
\begin{table*}
\centering
\caption{\protect\label{tab:Example-of-Two-Amps}Example of two amplifiers. Colored values indicate differences between the two devices. Metrics highlighted in red fail to distinguish the amplifiers, while measures highlighted in green ($M_{\eta}$ and $M_\text{W}$) clearly differentiate them and support correct ordering in a cascade.}
\begin{tabular}{c||c|c|c|c|c||c|c|c|c||c|c}
\hline
\rowcolor[HTML]{31668D}
{\footnotesize\textcolor{white}{Device}}
& {\footnotesize\textcolor{white}{$P_\text{in}$}}
& {\footnotesize\textcolor{white}{$G$}}
& {\footnotesize\textcolor{white}{$P_\text{dc}$}}
& {\footnotesize\textcolor{white}{$\text{IP}$}}
& {\footnotesize\textcolor{white}{$P_{\text{non-signal}}$}}
& {\footnotesize\textcolor{white}{$\eta$}}
& {\footnotesize\textcolor{white}{$W$}}
& {\footnotesize\textcolor{white}{$\eta^\text{max}$}}
& {\footnotesize\textcolor{white}{$W^\text{min}$}}
& {\footnotesize\textcolor{white}{\textbf{$M_{\eta}$}}}
& {\footnotesize\textcolor{white}{\textbf{$M_\text{W}$}}}
\tabularnewline
\hline\hline
{\footnotesize\textcolor{ampone}{Amp.\ \#1}}
& \cellcolor[HTML]{D6E8F5}{\footnotesize\textcolor{ampone}{1~$\mu$W}}
& \cellcolor[HTML]{D6E8F5}{\footnotesize 10}
& \cellcolor[HTML]{D6E8F5}{\footnotesize\textcolor{ampone}{1~mW}}
& \cellcolor[HTML]{D6E8F5}{\footnotesize\textcolor{ampone}{0.05~mW}}
& \cellcolor[HTML]{D6E8F5}{\footnotesize\textcolor{ampone}{0.991~mW}}
& \cellcolor[HTML]{FFD6D6}{\footnotesize\textcolor{red}{1\%}}
& \cellcolor[HTML]{FFD6D6}{\footnotesize\textcolor{red}{100.1}}
& \cellcolor[HTML]{FFD6D6}{\footnotesize\textcolor{red}{50\%}}
& \cellcolor[HTML]{FFD6D6}{\footnotesize\textcolor{red}{2.1}}
& \cellcolor[HTML]{EAF4D3}{\footnotesize\textbf{\textcolor{green!70!black}{0.00045}}}
& \cellcolor[HTML]{EAF4D3}{\footnotesize\textbf{\textcolor{green!70!black}{2222}}}
\tabularnewline
\hline
{\footnotesize\textcolor{amptwo}{Amp.\ \#2}}
& \cellcolor[HTML]{D6E8F5}{\footnotesize\textcolor{amptwo}{1~mW}}
& \cellcolor[HTML]{D6E8F5}{\footnotesize 10}
& \cellcolor[HTML]{D6E8F5}{\footnotesize\textcolor{amptwo}{1~W}}
& \cellcolor[HTML]{D6E8F5}{\footnotesize\textcolor{amptwo}{0.05~W}}
& \cellcolor[HTML]{D6E8F5}{\footnotesize\textcolor{amptwo}{0.991~W}}
& \cellcolor[HTML]{FFD6D6}{\footnotesize\textcolor{red}{1\%}}
& \cellcolor[HTML]{FFD6D6}{\footnotesize\textcolor{red}{100.1}}
& \cellcolor[HTML]{FFD6D6}{\footnotesize\textcolor{red}{50\%}}
& \cellcolor[HTML]{FFD6D6}{\footnotesize\textcolor{red}{2.1}}
& \cellcolor[HTML]{EAF4D3}{\footnotesize\textbf{\textcolor{green!70!black}{0.45}}}
& \cellcolor[HTML]{EAF4D3}{\footnotesize\textbf{\textcolor{green!70!black}{2.222}}}
\tabularnewline
\hline
\end{tabular}
\end{table*}
At their specified input powers, both devices exhibit identical values of $\eta$, $W$, $\eta^{\text{max}}$, and $W^{\text{min}}$. Nevertheless, Amp.~\#2 consumes nearly a watt of dc power, whereas Amp.~\#1 consumes only a milliwatt. This large disparity shows that $\eta$ and $W$—even when evaluated at their optimal operating points—do not reliably distinguish devices for system‑level design or determine their preferred ordering in a cascade.

Although $P_\text{dc}$ differentiates these two extreme examples, it is not generally a dependable indicator. Linearity constraints often limit the usable input power, causing two amplifiers with similar $\eta$ and $W$ to behave differently once realistic signal swing and EVM limits are imposed. Moreover, in a cascade, total dc power is independent of device ordering and therefore cannot guide architectural decisions.

This limitation mirrors the well‑known issue with noise factor: optimizing only $F$ encourages “through‑connection” solutions that minimize noise but provide no gain~\cite{Haus1958}. Analogously, optimizing purely for $\eta$ or $W$ favors devices that appear efficient at a narrow operating point but may be unusable over the required signal range.

The proposed efficiency and waste measures, $M_\eta$ and $M_W$ in~\eqref{eq:eta-measure}, address this by capturing the maximum input signal power each device can handle while meeting linearity and EVM requirements. When applied to Table~\ref{tab:Example-of-Two-Amps}, $M_\eta$ and $M_W$ clearly separate the two amplifiers, despite their identical $\eta$ and $W$ values. Both measures indicate that Amp.~\#1 should precede Amp.~\#2 in a cascade—consistent with the ordering predicted by the noise, linearity, and dynamic‑range measures discussed earlier, and consistent with practical system design.

\section*{\protect{\label{sec:simplified-examples}}Sidebar E: Simplified $\eta_\text{o}$ Expressions for Common Devices}

Although the definition of $\eta_\text{o}$ in \eqref{eq:SF-defined} may have complex dependence on $P_\text{in}$, closed-form expressions can be obtained for several common device models.

\subsection{\protect\label{subsec:Linear-amplifier-with-uniformPin}Linear Amplifier with Uniformly Distributed $P_\text{in}$}

For a linear amplifier with constant gain $G$, constant $P_\text{dc}$, sensitivity $P_\text{in}^\text{min}$, maximum input power $P_\text{in}^\text{max}$, and uniformly distributed $P_\text{in}$ between $P_\text{in}^\text{min}$ and $P_\text{in}^\text{max}$, \eqref{eq:SF-defined} simplifies to
\begin{align}
\eta_\text{o} & =G\times\frac{P_\text{in}^\text{max}+P_\text{in}^\text{min}}{2P_\text{dc}}\label{eq:SF-equal-prob}\\
 & =\frac{G\times P_\text{in}^\text{min}}{2P_\text{dc}}\left(\text{DRo}+1\right)\label{eq:SF-LNA}\\
 & =\frac{P_\text{out}^\text{max}}{2P_\text{dc}}\left(1+\text{DRo}^{-1}\right)\overset{\text{DRo}\gg1}{\approx}\frac{1}{2}\eta^\text{max},\label{eq:SF-PA}
\end{align}
where \eqref{eq:SF-LNA} is better suited to LNAs and \eqref{eq:SF-PA} to linear PAs. With respect to the data rate under the simplified assumptions in Section~\ref{sec:Signal-quality}, 
\eqref{eq:Data-rate-F-DR-IP-1} can be employed to recast \eqref{eq:SF-LNA}  and \eqref{eq:SF-PA} as
\begin{equation}
\begin{split}
\eta_\text{o}
&=\frac{G\times P_\text{in}^\text{min}}{2P_\text{dc}}
\left(\text{DR}_\text{max}\times\Gamma^{-3/2}\times2^{-3R/2B}+1\right)\\
&=\frac{P_\text{out}^\text{max}}{2P_\text{dc}}
\left(1+\text{DR}_\text{max}^{-1}\times\Gamma^{3/2}\times2^{3R/2B}\right)
\end{split}
\label{eq:SF-LNA-1-2}
\end{equation}
Equations in~\eqref{eq:SF-LNA-1-2} directly relate operating efficiency to channel capacity, bandwidth, and dynamic range, revealing the power cost of different design targets.

As an example, applying \eqref{eq:SF-equal-prob} to the measured ``fbgLNA'' of~\cite{Radpour2024}---a wideband CMOS LNA characterized experimentally at 10~GHz with $F \approx 2\,\text{dB}$, $P_\text{dc}=7.8\,\text{mW}$, $\text{IP}_{1\text{dB}}\approx-21\,\text{dBm}$, and $G\approx16\,\text{dB}$---yields $\eta_\text{o}^\text{LNA}\approx2\%$ under the assumptions of uniform $P_\text{in}$, $\text{SNR}=1$, and $B=0.1$~GHz. 

\subsection{Linear Amplifier with an M-QAM $P_\text{in}$}

For an M-QAM input signal with PDF
\begin{equation}
\mathscr{P}(P_\text{in}) = \frac{1}{M} \sum_{i=1}^{M} \delta\left(P_\text{in} - P_{\text{s,}i}\right), \label{eq:M-QAM-PDF}
\end{equation}
where $P_{\text{s,}i}$ denotes the power level of the $i$-th constellation symbol, the operating efficiency can be expressed as
\begin{equation}
\eta_\text{o} =\frac{G}{M\times P_\text{dc}}\sum_{i=1}^{M}P_{\text{s,}i} =G\times\frac{P_\text{in}^\text{ave}}{P_\text{dc}},
\label{eq:SF-M-QAM-lin-amp}
\end{equation}
where $P_\text{in}^\text{ave}$ denotes the average input power.

As an illustration, consider a PA operating with average backoff to avoid clipping at $P_\text{in}^\text{max}$. For a 64-QAM signal, $P_\text{out}^\text{min}$ corresponds to the lowest-power constellation point. The $i$-th symbol power is $P_{\text{s,}i} = 2Z^{2}A^{2}$, where $Z \in \{\pm1,\pm3,\pm5,\pm7\}$. For example, applying \eqref{eq:SF-M-QAM-lin-amp} to the measured class-J PA of~\cite{Rezaei2013}, which exhibits $\eta \approx 56\%$ at $P_\text{out}^\text{max} \approx 26\,\text{dBm}$ and $G \approx 9\,\text{dB}$ at 2.5\,GHz, and setting $A$ such that $\max(P_{\text{s,}i}) = P_\text{out}^\text{max}$ yields an estimated $\eta_\text{o}^\text{PA} = 24\%$.

\subsection{Doherty PA with Uniformly Distributed $P_\text{in}$}

In nonlinear PAs, $P_\text{dc}$ and $G$ vary with $P_\text{in}$. A useful approximation is to partition operation into regions where $P_\text{dc}$ and $G$ are approximately constant.

\begin{figure}
\centering
\includegraphics[width=1\columnwidth]{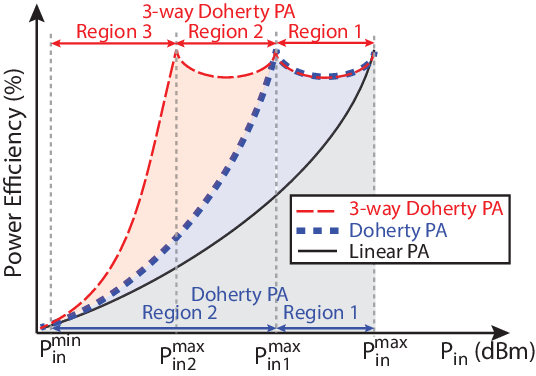}
\caption{\protect\label{fig:Sketch-of-efficiency}Sketch of power efficiency vs $P_\text{in}$ for a linear PA, a Doherty PA, and a 3-way
Doherty PA, exhibiting one and two power-backoff efficiency peaks, respectively. Region~2 of the Doherty PA spans Regions~2 and~3 of the 3-way Doherty PA, as the latter's additional peaking amplifier introduces an intermediate efficiency peak within that range.}
\end{figure}

Doherty PAs are designed to enhance efficiency at specific power levels.  Fig. \ref{fig:Sketch-of-efficiency} 
illustrates a typical efficiency curve for a Doherty PA and a 3-way Doherty PA with multiple power-efficiency-enhanced regions~\cite{Piacibello2023}. At each ``bump'' in power efficiency, the PA internal sub-circuits are either activated or deactivated, affecting both $P_\text{dc}$ and $G$. To account for these different regions, the integration range in \eqref{eq:SF-defined} is split into regions (as identified in Fig. \ref{fig:Sketch-of-efficiency}), within which $P_\text{dc}$ and $G$ remain constant, thereby allowing us to  express the operating efficiency of a Doherty PA as
\begin{equation}
\eta_\text{o}  =\underbrace{\int_{P_\text{in}^\text{min}}^{P_\text{in1}^\text{max}}\eta\left(P_\text{in}\right)\mathscr{P}\left(P_\text{in}\right)\text{d}P_\text{in}}_\text{Doherty PA: Region 2} +\underbrace{\int_{P_\text{in1}^\text{max}}^{P_\text{in}^\text{max}}\eta\left(P_\text{in}\right)\mathscr{P}\left(P_\text{in}\right)\text{d}P_\text{in}}_\text{Doherty PA: Region 1}\label{eq:SF-definition-doherty}
\end{equation}
which, for a uniformly distributed $P_\text{in}$, reduces to
\begin{align}
\eta_\text{o} & =\frac{\left(P_\text{in1}^\text{max}\right)^{2}}{P_\text{in}^\text{max}-P_\text{in}^\text{min}}\left[G_\text{r1}\frac{1-\text{DRo}_\text{r1}^{-2}}{2P_\text{dc,r1}}+G_\text{r2}\frac{\text{DRo}_\text{r2}^{2}-1}{2P_\text{dc,r2}}\right]
\end{align}
where $\text{DRo}_\text{r1}=P_\text{in1}^\text{max}/P_\text{in}^\text{min}$, $\text{DRo}_\text{r2}=P_\text{in}^\text{max}/P_\text{in1}^\text{max}$, and $P_\text{dc,r1(2)}$ and $G_\text{r1(2)}$ are the power consumption and
gains in Doherty regions \#1 and \#2, respectively.

\subsection{3-Way Doherty PA with an M-QAM Input Signal}
The integral in~\eqref{eq:SF-definition-doherty} generalizes naturally to $N$ regions by adding further integration intervals, as required for, for example, a 3-way Doherty PA.
For an M-QAM input signal with PDF \eqref{eq:M-QAM-PDF}, the operating efficiency of a 3-way Doherty PA is

\begin{equation}
\eta_\text{o}\approx\frac{1}{M}\left[\frac{G_\text{r1}}{P_\text{dc,r1}}\sum_{i=1}^{M_\text{bo}}P_{\text{s,}i}+\frac{G_\text{r2}}{P_\text{dc,r2}}\sum_{i=M_\text{bo}+1}^{M}P_{\text{s,}i}\right],
\end{equation}
where $i=1..M_\text{bo}$ identifies constellation points with power less than $P_\text{in1}^\text{max}$, and $i>M_\text{bo}$ is for the other constellation points.

As an illustration, the 28-GHz Doherty PA of~\cite{HuaWang2024} achieves $\eta = 39\%$ at $P_\text{out}^\text{max} = 20.4\,\text{dBm}$, $\eta = 30\%$ at $P_{\text{out1}}^\text{max} = 14.4\,\text{dBm}$ (6-dB backoff for 64-QAM), $\eta = 20.5\%$ at $P_{\text{out2}}^\text{max} = 11.4\,\text{dBm}$ (9-dB backoff for 256-QAM), and $\eta \approx 3\%$ at $P_\text{out}^\text{min} = 3.4\,\text{dBm}$. The low-power gain is 18~dB, decreasing to about 16.7~dB at $P_\text{out}^\text{max}$. Evaluating~\eqref{eq:SF-defined} across the three regions identified in Fig.~\ref{fig:Sketch-of-efficiency}, with $G$ and $P_\text{dc}$ assumed constant within each region, and considering 64-QAM and 256-QAM modulation schemes separately, yields $\eta_\text{o}^{\text{64-QAM}} = 21\% \text{(Region 1)} + 6\% \text{(Region 2)}+ 2\% \text{(Region 3)}= 29\%$ and $\eta_\text{o}^{\text{256-QAM}} = 19\% \text{(Region 1)}+ 6\% \text{(Region 2)}+ 1\% \text{(Region 3)}= 26\%$.

\section*{Acknowledgment}

The authors would like to acknowledge the financial support of the Natural Sciences and Engineering Research Council of Canada (NSERC), the National Science Foundation (NSF), the Office of Naval Research (ONR), U.S. Department of Energy, Office of Science, Office of Advanced Scientific Computing Research, Early Career Research Program under Award Number DE-SC0023957, and CMC Microsystems.

\ifCLASSOPTIONcaptionsoff
  \newpage
\fi

%
%
%
\bibliographystyle{IEEEtran}
\bibliography{bibliography}
%
%


\vfill


\end{document}